# BINARY INTERACTION WITH MULTIPLE FLUID TYPE COSMOLOGY UNDER MODIFIED GRAVITY FRAME

Alokananda Kar[1] and Shouvik Sadhukhan[2]

1. Department of Physics; University of Calcutta; 92 APC Road, Kolkata 700009, West Bengal, India
2. Department of Physics; Indian Institute of Technology, Kharagpur 721302, West Bengal, India

Corresponding Author's Email: alokanandakar@gmail.com

## ABSTRACT

In the present chapter, we have established multiple fluid cosmological models under interaction scenarios. The interaction model we have established is a binary type of interaction scenario where three types of fluids are bound with interaction. We have incorporated variable gravitational constant and variable cosmological constant. The whole work has proceeded with modified gravity geometry where the modified effect over Einstein's gravity acts like a variable cosmological constant. We have used linear dark matter (DM), inhomogeneous fluid and a new type of non-linear model as interacting components. We have discussed the cosmic phases only with fluid dynamical approaches i.e., through the scale factor variations of effective energy density, pressure, equation of state (EOS) parameter and gravitational constant ($\Lambda$).

## 1. INTRODUCTION

Recent observational studies proved that the universe should be in the accelerated expanding stage [Capozziello.et.al (2002,2003 and 2006)]. The observational studies developments brought several problems in front of researchers viz. Cosmological horizon problem, Magnetic monopole problem, and Vacuum energy problem. For the resolution of these problems, the idea of the exotic nature of cosmic matter has been brought into physics [Kar.et.al (2020-2022) and Sadhukhan.et.al (2020)]. Interacting Dark matter, Non-linear fluid cosmology are among many examples of exotic matters. [Chattopadhyay.et.al (2009,2010 and 2017)]

Dark matter is one candidate that discusses such an expansion scheme of the universe but for the non-relativistic case, they can't. The non-relativistic dark matter is considered cold dark matter and they produce zero pressure in large-scale structures [Debnath.et.al (2008 and 2013)]. The warm dark matter or hot dark matter can produce large positive energy, which is again considered relativistic dark matter and theoretically, relativistic cosmic matter can dominate the universe only after certain initial conditions or causes [Dirac.et.al (1974 and 1979)]. Thus, the implementation of non-exotic matter or dark matter can't solve the problems regarding accelerated expanding cosmology instead brings different problems. Hence, exotic type of cosmic matter or exotic dark matter has come into play [Hoyle.et.al (1964 and 1971)].

The exotic dark matter acts like the Cardassian model where the exotic nature comes from the self-interaction produced between dark matters [Zeldovich.et.al (1968)]. This self-interaction brings the non-linearity and inhomogeneity into the equation of the state of dark matter. Hence, inhomogeneous, and non-linear models came into play [Banerjee.et.al (1985)]. Those inhomogeneous and non-linear models are sometimes called fluid-type dark energy models. Another important cause behind the introduction of the inhomogeneous model is to provide a pause of cosmic inflation and bring the reheating phase a start in late time accelerated expanding phases. Chaplygin gas, and Van-Der-Waals fluid are two examples of non-linear models [Bergmann.et.al (1968)].

The modified gravity is an alternative geometry analysis mechanism which can substitute the non-linear and inhomogeneous type self-interactions to discuss the accelerated expansion of the universe. The modified gravity provides the modifications in the Einstein Hilbert action with some higher ordered terms. $f(R), f(G), f(T)$ and $f(Q)$ gravities are the common forms of modified gravity theory. In our present work, we have given the derivations of $f(R,T)$ gravity with the functional form of $f(R,T) = R + 2\lambda T$ [Nojiri.et.al (2006)] Introduction of variable gravitational constant and cosmological constant and their coupling during the dissociation of the divergence of energy momentum tensor, can also be considered as an alternative to the modified gravity which can produce accelerated expanding universe model with the variation of $G$ and $\Lambda$. The variable gravitational constant can give the variation of expansion scheme during the phase transitions. The variable cosmological constant varies the repulsive effect on cosmic bubbles due to $\Lambda-$type dark energies [Brevik.et.al (2007)].

In our present chapter, we have established interaction between three types of fluid systems viz. linear dark matter, inhomogeneous fluid, and a new type of three parameter-based non-linear fluid. The three fluids discussed the binary interaction scenario. The interaction picture provided the functional variations of gravitational constant and different cosmic phases. We have included the mathematical analysis only [Kar.et.al (2020-2022)].

The chapter has proceeded according to; in section 2 we have given the theoretical mechanism of the whole fluid dynamical process including the interaction scenario and the analysis of the mathematical results. Finally, in section 3 we have concluded the chapter.

## 2. THEORETICAL FORMALISM OF THE MODEL

The theoretical and mathematical mechanism of our present chapter started with the following action for the multiple fluid cosmologies. Here, the cosmic matter Lagrangian is a coupled form of linear dark matter, inhomogeneous fluid, and a new type of non-linear fluid system. The system works on Modified gravity geometry and the presence of variable gravitational constant. Hence, the action can be written as follows.

$$S = \frac{1}{16\pi} \int d^4x \sqrt{-g} \left( \frac{f(R,T)}{G} + L_m^{eff} \right) \qquad (1)$$

After applying the least action principle to equation (1) we can find the following dynamical equation of cosmic dynamics. The functional form we considered for modified gravity is $f(R,T) = R + 2\lambda T$ with $T = Tr[T_{\mu\nu}] = \rho_{eff} - 3p_{eff}$. Hence, we have considered $\Lambda = \lambda(\rho_{eff} - p_{eff})$. Thus, we obtain the following equation.

$$G_{\mu\nu} = 8\pi G(t) T_{\mu\nu}^{eff} + \Lambda g_{\mu\nu} \qquad (2)$$

Here we can write $T_{\mu\nu}^{eff} = -\frac{2}{\sqrt{-g}} \frac{\delta(\sqrt{-g} L_m^{eff})}{\delta g^{\mu\nu}}$ and $G_{\mu\nu} = R_{\mu\nu} - \frac{1}{2} R g_{\mu\nu}$. Now the symmetry and other geometric modifications due to application or coupling of cosmic matters with geometry can be discussed using the line element of the system and the evolution of the cosmic dynamical equation given in equation (3). Hence, we can consider the following line element.

$$ds^2 = dt^2 - a^2(t)(dr^2 + r^2 d\Omega^2) \qquad (3)$$

We have considered the flat FRW geometry with scale factor $a(t)$ and angular coordinate $d\Omega^2 = d\theta^2 + \sin^2(\theta) d\phi^2$. Using this line element into the Einstein equation discussed in equation (3) we can find the following two equations of cosmic dynamics i.e., the FRW equations. Hence, we can write as follows.

$$3H^2 = 8\pi G \rho_{eff} + \Lambda \tag{4a}$$

And,

$$3H^2 + 2\dot{H} = -8\pi G p_{eff} + \Lambda \tag{4b}$$

Now for each cosmological model, we should nullify the divergence of the Einstein tensor as well as the energy-momentum tensor. The divergence-less condition of energy-momentum tensor can provide the idea of conservation of energy-momentum for an effective system of cosmic fluids under cosmic horizons. Using the divergence-less condition of Einstein tensor as discussed in equation (2), we can find the following equation.

$$8\pi \dot{G} \rho_{eff} + \dot{\Lambda} + 8\pi G \left( \dot{\rho}_{eff} + 3H(\rho_{eff} + p_{eff}) \right) = 0 \tag{5}$$

Now from the divergence-less property of the energy-momentum tensor of the effective fluid system and the equation (5), we can find the following two equations.

$$\dot{\rho}_{eff} + 3H(\rho_{eff} + p_{eff}) = 0 \tag{6a}$$

And,

$$8\pi \dot{G} \rho_{eff} + \dot{\Lambda} = 0 \tag{6b}$$

Hence, we can find,

$$G(a) = G_0 - \int \frac{\Lambda(a)}{8\pi \rho_{eff}(a)} da \tag{7}$$

For the cosmological constant, we assumed it to be variable like the relation $\Lambda = \alpha H^2$. Hence, we can find the following differential equation for the time variation solution of the scale factor.

$$2a(t)\ddot{a}(t) + 4\dot{a}^2(t) + 8\pi G(a)\left(p_{eff} - \rho_{eff}\right)a^2(t) - 2\alpha H_0^2 a^{2(\beta+1)} = 0 \tag{8}$$

In the interaction scenarios, we consider the equation of states as $p_m = \omega_m \rho_m$, $p_f = A\rho_f + BH^2$ and $p_{Nf} = C\rho_{Nf} + D\rho_{Nf}^2 - \frac{D}{\rho_{Nf}^{\alpha_1}}$. Considering this equation of states in multiple fluid cosmology, we can expand the energy momentum conservation equation (6a) as follows.

$$\dot{\rho}_m + 3H(\rho_m + p_m) = -3H\rho_m \delta_1 - 3H\rho_m \delta_2 \tag{9a}$$

$$\dot{\rho}_f + 3H(\rho_f + p_f) = 3H\rho_m \delta_1 \tag{9b}$$

And,

$$\dot{\rho}_{Nf} + 3H(\rho_{Nf} + p_{Nf}) = 3H\rho_m \delta_2 \tag{9c}$$

The solutions for energy densities and pressures can be given in table II. For the derivations of the gravitational constant, we found the same conditions as follows in table I.

**TABLE I: List of Conditions for this model**

| **Primary Conditions:** | • Dark matter dominant: $\rho_m \gg \rho_f \gg \rho_{Nf}$ <br> • Inhomogeneous fluid dominant: $\rho_f \gg \rho_m \gg \rho_{Nf}$ <br> • Non-linear fluid dominant: $\rho_{Nf} \gg \rho_f \gg \rho_m$ <br> • Mixture dominant: $\rho_m \approx \rho_f \approx \rho_{Nf}$ |
|---|---|
| **Secondary Conditions:** | • Mixture dominant Case with very large Powers: $-3(1+\omega_m) \approx -3(1+A) \approx 1 + 2\beta \gg a^{\frac{3}{\sqrt{4D^2 E + 1}}} \gg 1$ <br> • Mixture dominant Case with small Powers: $-3(1+\omega_m) \approx -3(1+A) \approx 1 + 2\beta \approx a^{\frac{3}{\sqrt{4D^2 E + 1}}} \gg 1$ |

The functional forms of the gravitational constant for the two secondary conditions can be given as follows according to the order of the conditions list.

$$G(a) = G_0 - \frac{\lambda}{8\pi}\ln(\rho_m + \rho_f + \rho_{Nf}) + \frac{\lambda}{8\pi} * Constant * \ln(a) \tag{10a}$$

And,

$$G(a) = G_0 - \frac{\lambda}{8\pi}\ln(\rho_m + \rho_f + \rho_{Nf}) + \frac{\lambda}{8\pi} * Constant_1 * \ln(a) \qquad (10b)$$

Here, the terms $Constant$ and $Constant_1$ are as follows.

$$Constant = \frac{3(\delta_1 + \delta_2 - \omega_m)(1+\omega_m)\rho_{m0} - 3(1+A)A\rho_{f0} + \frac{2(1+2\beta)(2+\beta)B}{3A+2\beta+4} - 3(1+\omega_m)(\delta_1+\delta_2)\rho_{m0}}{\rho_{m0} + \rho_{f0} - \frac{3B}{3A+2\beta+4}}$$

And,

$$Constant_1 = \frac{3(\delta_1+\delta_2-\omega_m)(1+\omega_m)\rho_{m0} - 3(1+A)A\rho_{f0} + \frac{2(1+2\beta)(2+\beta)B}{3A+2\beta+4} - 3(1+\omega_m)(\delta_1+\delta_2)\rho_{m0} - \frac{1}{2D}\sqrt{4D^2E+1}\, c_1 c_2}{\rho_{m0}+\rho_{f0}-\frac{3B}{3A+2\beta+4}-\frac{c_1}{D}\sqrt{4D^2E+1}}$$

(11)

**TABLE II: Energy densities and pressures for triple fluid interaction scenario in modified geometry**

| Name of fluid | Energy density | Pressure |
|---|---|---|
| Linear dark Matter | $\rho_m = \rho_{m0}a^{-3(1+\omega_m)}$ | $p_m = (\omega_m - \delta_1 - \delta_2)\rho_{m0}a^{-3(1+\omega_m)}$ |
| Inhomogeneous fluid | $\rho_f = \rho_{f0}a^{-3(1+A)} - \frac{3B}{3A+2\beta+4}a^{1+2\beta}$ | $p_f = \delta_1\rho_{m0}a^{-3(1+\omega_m)} + (A)\rho_{f0}a^{-3(1+A)} + \frac{B(2\beta+4)}{3A+2\beta+4}a^{2\beta+1}$ |
| New Non-linear fluid | $\rho_{Nf} = \sqrt{E + \frac{1}{4D^2}} \left( \frac{\left(\frac{\rho_{Nf0}+\frac{1}{2D}+\sqrt{E+\frac{1}{4D^2}}}{\rho_{Nf0}+\frac{1}{2D}-\sqrt{E+\frac{1}{4D^2}}}\right) a^{\frac{3}{\sqrt{4D^2E+1}}}+1}{\left(\frac{\rho_{Nf0}+\frac{1}{2D}+\sqrt{E+\frac{1}{4D^2}}}{\rho_{Nf0}+\frac{1}{2D}-\sqrt{E+\frac{1}{4D^2}}}\right) a^{\frac{3}{\sqrt{4D^2E+1}}}-1} \right)$ | $p_{Nf} = \delta_2\left(\rho_{m0}a^{-3(1+\omega_m)}\right) - \left( \sqrt{E+\frac{1}{4D^2}} \left( \frac{\left(\frac{\rho_{Nf0}+\frac{1}{2D}+\sqrt{E+\frac{1}{4D^2}}}{\rho_{Nf0}+\frac{1}{2D}-\sqrt{E+\frac{1}{4D^2}}}\right) a^{\frac{3}{\sqrt{4D^2E+1}}}+1}{\left(\frac{\rho_{Nf0}+\frac{1}{2D}+\sqrt{E+\frac{1}{4D^2}}}{\rho_{Nf0}+\frac{1}{2D}-\sqrt{E+\frac{1}{4D^2}}}\right) a^{\frac{3}{\sqrt{4D^2E+1}}}-1} \right) \right) - \frac{1}{2D} - \frac{1}{D}\left( \frac{\left(\frac{\rho_{Nf0}+\frac{1}{2D}+\sqrt{E+\frac{1}{4D^2}}}{\rho_{Nf0}+\frac{1}{2D}-\sqrt{E+\frac{1}{4D^2}}}\right) a^{\frac{3}{\sqrt{4D^2E+1}}}}{\left(\frac{\rho_{Nf0}+\frac{1}{2D}+\sqrt{E+\frac{1}{4D^2}}}{\rho_{Nf0}+\frac{1}{2D}-\sqrt{E+\frac{1}{4D^2}}}\right) a^{\frac{3}{\sqrt{4D^2E+1}}}-1} \right)^2$ |

The equation of state parameter can be written using the above relations of pressures and densities. Hence, the $\omega_{eff}$ can be written as follows.

$$\omega_{eff} = \left( (\omega_m - \delta_1 - \delta_2)\rho_{m0}a^{-3(1+\omega_m)} + \delta_1\rho_{m0}a^{-3(1+\omega_m)} + (A)\rho_{f0}a^{-3(1+A)} + \frac{B(2\beta+4)}{3A+2\beta+4}a^{2\beta+1} + \delta_2\left(\rho_{m0}a^{-3(1+\omega_m)}\right) - \right.$$

$$\left. \left( \sqrt{E + \frac{1}{4D^2}} \left( \frac{\left(\frac{\rho_{Nf0}+\frac{1}{2D}+\sqrt{E+\frac{1}{4D^2}}}{\rho_{Nf0}+\frac{1}{2D}-\sqrt{E+\frac{1}{4D^2}}}\right)a^{\frac{3}{\sqrt{4D^2E+1}}}+1}{\left(\frac{\rho_{Nf0}+\frac{1}{2D}+\sqrt{E+\frac{1}{4D^2}}}{\rho_{Nf0}+\frac{1}{2D}-\sqrt{E+\frac{1}{4D^2}}}\right)a^{\frac{3}{\sqrt{4D^2E+1}}}-1} \right) - \frac{1}{2D} - \frac{1}{D}\left( \frac{\left(\frac{\rho_{Nf0}+\frac{1}{2D}+\sqrt{E+\frac{1}{4D^2}}}{\rho_{Nf0}+\frac{1}{2D}-\sqrt{E+\frac{1}{4D^2}}}\right)a^{\frac{3}{\sqrt{4D^2E+1}}}}{\left(\frac{\rho_{Nf0}+\frac{1}{2D}+\sqrt{E+\frac{1}{4D^2}}}{\rho_{Nf0}+\frac{1}{2D}-\sqrt{E+\frac{1}{4D^2}}}\right)a^{\frac{3}{\sqrt{4D^2E+1}}}-1} \right)^2 \right) \right) / \left( \rho_{m0}a^{-3(1+\omega_m)} + \rho_{f0}a^{-3(1+A)} - \frac{3B}{3A+2\beta+4}a^{1+2\beta} + \right.$$

$$\left. \sqrt{E + \frac{1}{4D^2}} \left( \frac{\left(\frac{\rho_{Nf0}+\frac{1}{2D}+\sqrt{E+\frac{1}{4D^2}}}{\rho_{Nf0}+\frac{1}{2D}-\sqrt{E+\frac{1}{4D^2}}}\right)a^{\frac{3}{\sqrt{4D^2E+1}}}+1}{\left(\frac{\rho_{Nf0}+\frac{1}{2D}+\sqrt{E+\frac{1}{4D^2}}}{\rho_{Nf0}+\frac{1}{2D}-\sqrt{E+\frac{1}{4D^2}}}\right)a^{\frac{3}{\sqrt{4D^2E+1}}}-1} \right) \right) \tag{12}$$

## 3. CONCLUSION

The chapter contains the mathematical analysis of multiple fluid cosmologies where we have tried to develop different equations of states through the coupled actions for multiple fluid models with interaction scenarios. The linear dark matter, inhomogeneous fluid and new type of non-linear fluids have been used here under modified geometry. The idea of variable gravitational constant and cosmological constant has also been interpreted into cosmic dynamics. The equation of state contains six dynamical parameters and one power parameter which can explain cosmological phases like Phantom phase, Quintessence, CDM, WDM and radiation as well as a smooth transition between phases like Quintom, Graceful exit etc.

## 4. FUTURE WORKS AND POSSIBILITIES

The equations of states derived here will be used in the detailed thermodynamic analysis in near future. This model will also be used to discuss the fifth force formalism as well as brane cosmic interpretations. The modified gravity to cosmological constant transition in action used here, will be used in detail to find the suitable entropies for the generalized second law of thermodynamics for the model.